\documentclass[slac_one]{revtex4}
\usepackage{graphicx}
\usepackage{fancyhdr}
\begin{document}
\title{The Production of Neutral Bilepton Bosons in Proton-Proton Collisions }
\author{E. Ramirez Barreto\footnote{elmer@if.ufrj.br}, Y. A. Coutinho \footnote{yara@if.ufrj.br}}
\affiliation{UFRJ, Rio de Janeiro, RJ 21941-972, Brazil}
\author{ J. S\'a Borges\footnote{saborges@uerj.br}}
\affiliation{UERJ, Rio de Janeiro, RJ 20550-013, Brazil}
\begin{abstract}
We establish some signatures of the extra bilepton boson ${X^0}$ predicted in the ${SU (3)_C \times SU (3)_L \times  U (1)_X }$ model with right-handed neutrinos. We analyze the process $p + p \longrightarrow X^0 +X^{0*} + {\hbox {anything}}$,
for center of mass energy regime of the Large Hadron Collider. The  main contributions for the neutral bilepton production in $q \bar q$  process come from the $s-$channel ($Z$ and $Z^\prime$ exchanges), when the initial  quarks have  charge $\frac{2}{3}$, and from an additional $t-$channel (heavy quark exchange) when they have  -$\frac{1}{3}$ of the positron electric charge.
We calculate some distributions of the final bileptons and from these results we conclude that LHC can show a clear signature for the existence of the $X^0$ predicted in the 3-3-1 model.
\end{abstract}
\maketitle
\thispagestyle{fancy}
\section{INTRODUCTION}
In contrast with the Standard Model (SM), the exotic gauge particles such as  leptoquarks and bileptons, predicted in many extensions of the SM, carry global quantum numbers. The bileptons are defined as bosons carrying two units of lepton number and are present in  the $3-3-1$ model.

The version of the $3-3-1$ model with right-handed neutrinos \cite{RHN} predicts the existence of  single-charged and neutral bileptons. The family  lepton number violation offers  a clear  signature for the bileptons detection \cite{DION}. The possibility of detecting neutral bileptons in future colliders  has not been  studied in the literature. 

In the present work we obtain the partial and the total width of the neutral bilepton $X^0$,  two elementary distributions, the total cross section and some final bilepton distributions for LHC energy regime.

\section{MODEL}
This version of the model includes  right-handed neutrinos for each leptonic family and 
two quark generations ($m=1,2$)  belong to anti-triplet and the other to triplet representation
\begin{eqnarray}
Q_{m L} = && \left(d^\prime_{m},\ -u^\prime_{m},\  D^\prime_{m} \right)_L^T \sim ({\bf 3}, {\bf 3^*}, 0),
\nonumber \\
Q_{3L} = && \left( t^\prime, \ b^\prime, \ T^\prime \right)_L^T \sim ({\bf 3}, {\bf 3}, 1/3),
\end{eqnarray}

The new quarks $D_{1}$ and $D_{2}$ carry $-\frac{1}{3}$ and  $T$,  $\frac{2}{3}$ units of positron charge. The gauge sector has an extra  neutral boson ($Z^\prime$) and bileptons ($V^\pm$ and $X^0$).

The important relation, used in the present work, between $Z^\prime$ and the bileptons ($V^\pm$ and $X^0$) masses is:
\begin{eqnarray}
{M_{V}\over M_{{Z^{\prime}}}}\simeq {M_{X}\over M_{{Z^{\prime}}}}\simeq{{\sqrt{3-4\sin^2\theta_W}}\over {2\cos\theta_W}}, 
\end{eqnarray}

The neutral current interactions involving quarks and  $Z$ or $Z^{\prime}$ is given by 
\begin{eqnarray}
{\cal L} = -  \frac{g}{2\cos\theta_W}\sum_f\bigl\lbrace \bar
\Psi_f\, \gamma^\mu\ (g_{v f} - g_{a f}\gamma^5)\ \Psi_f\, Z_\mu
 + \bar
\Psi_f\, \gamma^\mu\ (g^\prime_{v f} - g^\prime_{a f}\gamma^5)\ \Psi_f\, { Z_\mu^\prime}
\bigr\rbrace,
\end{eqnarray}
where the couplings $g^\prime_v$ and $g^\prime_a$ can be found in \cite{EYB}.

The Lagrangian for the interaction between the quarks  and the $X^0$ is:
\begin{eqnarray}
{\cal L} =  && -\frac{g}{2\sqrt{2}}\lbrace \bar t_{L} \gamma^\mu\ T_{L} - \bar D_{m L}\gamma^\mu\ d_{m L} \rbrace X^0 _\mu  + H.c.,
\end{eqnarray}
\section{ RESULTS}
 Let us consider first the partial and total width of the $X^0$. The main decay modes are $\bar q Q$ ($Q$ carries two units of lepton number) and $\nu \nu$. Fixing the  exotic quark masses equal to  $600$ GeV, we obtain the following values for $\Gamma_{X^0}$: $1.54$, $6.7$ and $14.7$ GeV for $M_{X^0}= 800 $, $1000$ and $1200$ GeV respectively.

The main contributions for $X^0$ pair production in $p \, p$  collision depend on the initial quark $q$  charge.  
When  $q = u$,  only $Z$ and $ Z^\prime$  {\it via} s-channel contribute and, on the other hand, when  $q = d$ we have an additional t-channel heavy quark exchange contribution. In our calculations we employed the  CompHep package \cite{COMP}.
Beginning with the dominant elementary  process $u \bar u$ and $d \bar d$, we find some final bileptons distributions. In the  Figure 1, for example,  we display  the $X^0$ angular distribution relative to the initial beam direction.  These distributions were  calculated by imposing the cuts on the final states

\centerline{$ -0.95 \leq\cos \theta_{1i} \leq 0.95, \,\,  -2.5 \leq y_{i}  \leq 2.5,  \,\,  p_{t} \geq 50$ GeV.}
We note that the angular distribution  shapes are different for $u$ and $d$ initial quarks. This is expected because $d \bar d$ receives an additional heavy quark $t-$channel contribution, leading to a more asymmetric distribution.

\begin{figure*}[t]
\centering
\includegraphics[width=75mm]{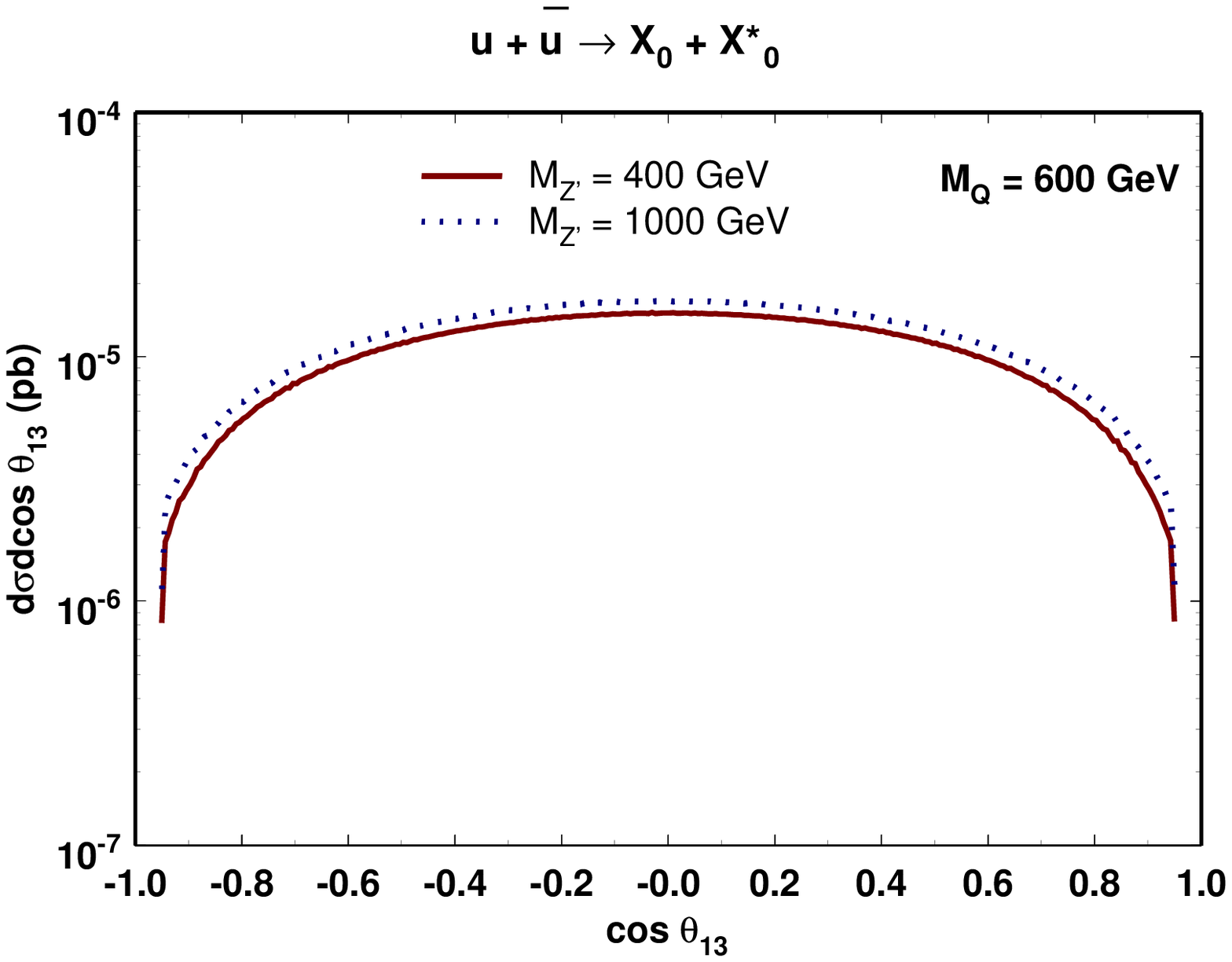}
\includegraphics[width=75mm]{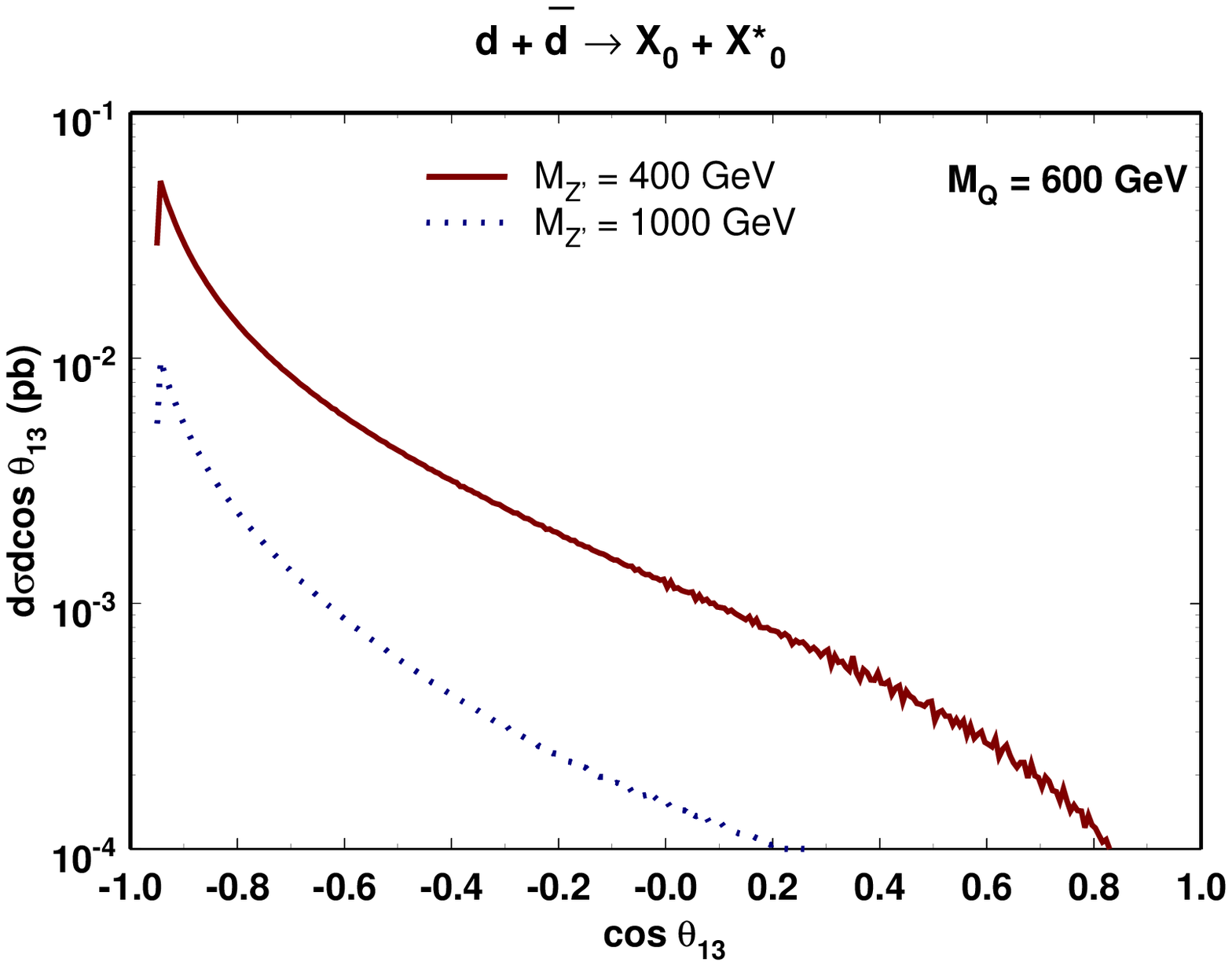}
\caption{Final bilepton $X^0$ angular distribution relative to the initial beam direction, considering $u \bar u$
 channel (left) and $d \bar d$ (right) for two different $Z^\prime$ masses.}
\end{figure*}
\begin{figure*}[t]
\centering
\includegraphics[width=75mm]{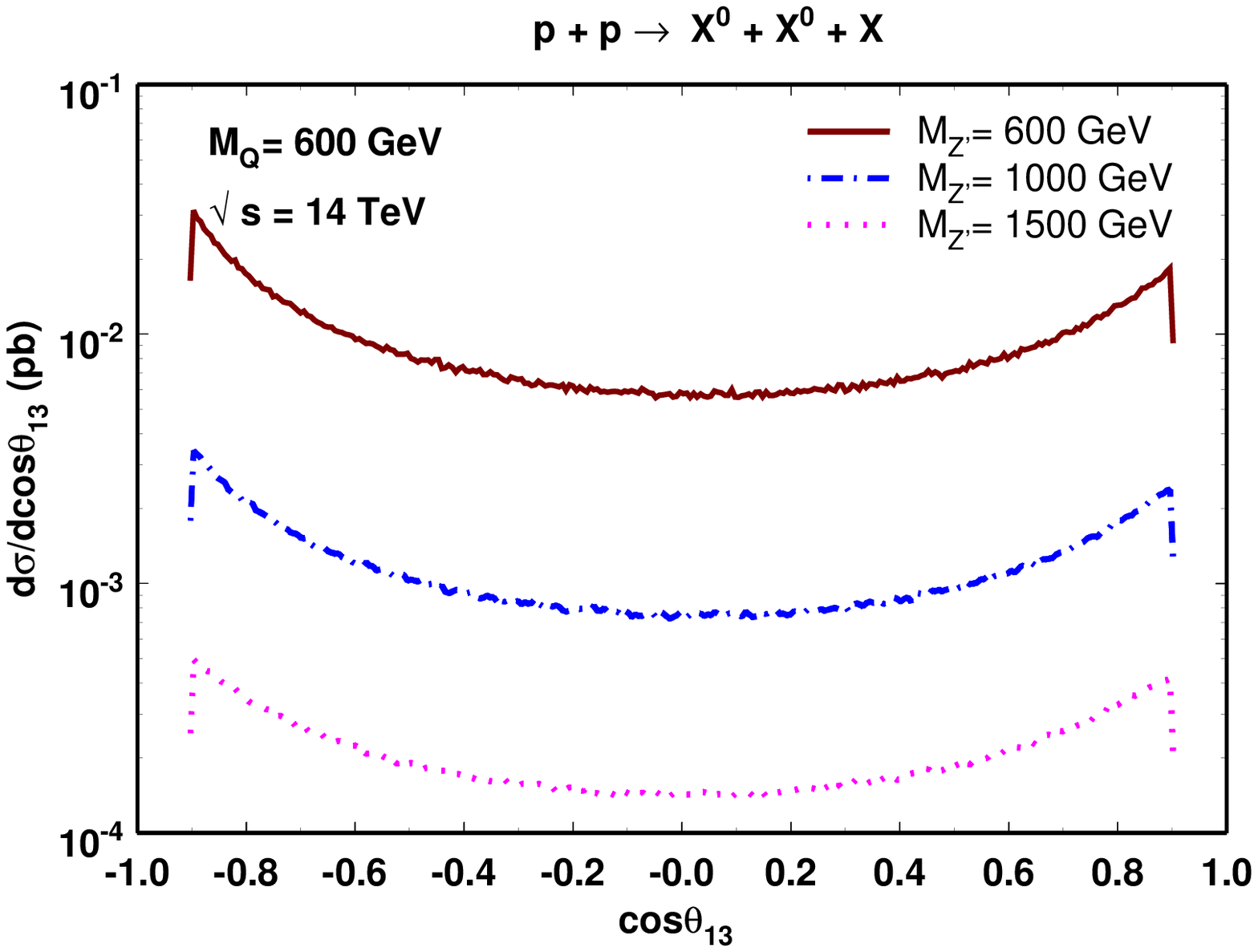}
\includegraphics[width=75mm]{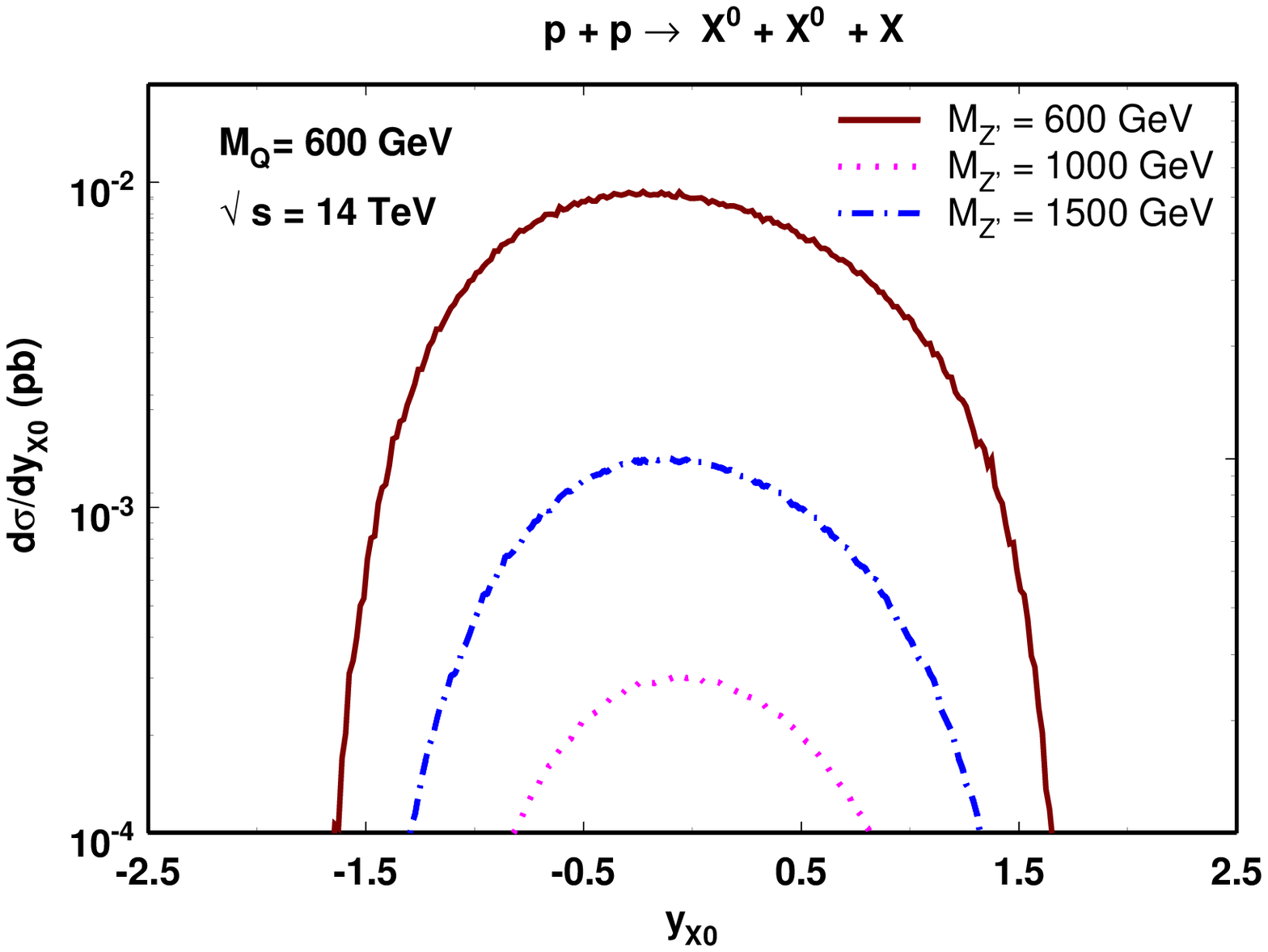}
\caption{Final bilepton $X^{0}$ angular distribution relative to the initial beam direction (left) and $X^0$ rapidity distribution (right), for three  different $Z^\prime$ masses}
\end{figure*}

\begin{figure*}
\centering
\includegraphics[width=75mm]{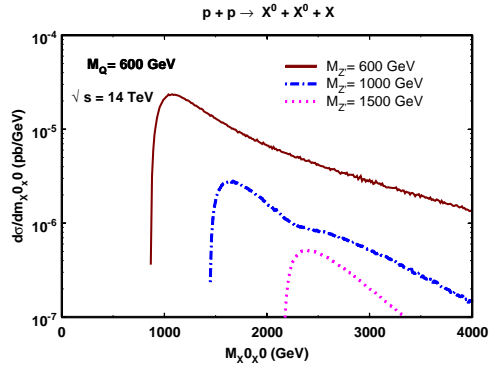}
\caption{Differential cross section vs. $ X^{0}$ pair invariant mass considering three different $Z^\prime$ masses.}
\end{figure*}

Now for $ p p$ collisions at $\sqrt s = 14$ TeV, we employ the CTEQ661 \cite{CTQ} structure functions. We show in Figure 2 the resulting $X^0$ angular distribution relative to the initial beam direction and the $X^0$ rapidity distribution.
The angular distribution shape is almost flat and are independent of $Z^{\prime}$ mass, while the $X^0$ rapidity distribution
is more concentrate for higher $Z^{\prime}$ masses.
Finally, we show in the Figure 3 the differential cross section as a function of the invariant $X^0$ pair mass. We note that the this distribution decreases when $M_{Z^{\prime}}$ increases.
\section{CONCLUSION}
In this work we have presented some initial results for the production of neutral bileptons predicted in the version of the 3-3-1 model with right-handed neutrinos. We have  shown  some distributions for the elementary process in $pp$ collisions. The inclusion of heavy quarks leads to unique signature and the $X^0$ decay into two leptons, via bileptoquark and charged bilepton.  For an annual luminosity at the LHC (${\cal L} = 100$\, fb$^{-1}$) we find $\simeq 100$ $X^0$ pairs produced {\it per} year. 
The next step in our analysis  is to consider the $X^0$ decay into leptons in order to compare it with $Z$ decay. This comparison can possibly reveal a signature of the neutral bilepton production.
\begin{acknowledgments}
E. Ramirez Barreto  thanks Capes and Y. A. Coutinho thanks FAPERJ for financial  support.
\end{acknowledgments}

\end{document}